\documentclass[reprint, aps, prstab, showpacs, showkeys, superscriptaddress]{revtex4-1}

\usepackage{graphicx}
\usepackage{dcolumn}
\usepackage{bm}

\begin{document} 

\title{Monochromaticity of coherent Smith-Purcell radiation from finite size grating}

\author{A.~Aryshev}
\email[Corresponding author: ]{alar@post.kek.jp}
\affiliation{KEK: High Energy Accelerator Research Organization, 1-1 Oho, Tsukuba, Ibaraki 305-0801, Japan}
\author{A.~Potylitsyn}
\affiliation{Tomsk Polytechnic University, Institute of Physics and Technology, Lenin Avenue 30, Tomsk 634050, Russian Federation}
\author{G.~Naumenko}
\email[Corresponding author: ]{naumenko@tpu.ru}
\affiliation{Tomsk Polytechnic University, Institute of Physics and Technology, Lenin Avenue 30, Tomsk 634050, Russian Federation}
\author{M.~Shevelev}
\affiliation{KEK: High Energy Accelerator Research Organization, 1-1 Oho, Tsukuba, Ibaraki 305-0801, Japan}
\author{K.~Lekomtsev}
\affiliation{KEK: High Energy Accelerator Research Organization, 1-1 Oho, Tsukuba, Ibaraki 305-0801, Japan}
\affiliation{John Adams Institute at Royal Holloway, University of London, Egham,  Surrey TW20 0EX, United Kingdom}
\author{L.~Sukhikh}
\affiliation{Tomsk Polytechnic University, Institute of Physics and Technology, Lenin Avenue 30, Tomsk 634050, Russian Federation}
\author{P.~Karataev}
\affiliation{John Adams Institute at Royal Holloway, University of London, Egham,  Surrey TW20 0EX, United Kingdom}
\author{Y.~Honda}
\affiliation{KEK: High Energy Accelerator Research Organization, 1-1 Oho, Tsukuba, Ibaraki 305-0801, Japan}
\affiliation{SOKENDAI: The Graduate University for Advanced Studies, 1-1, Oho, Tsukuba, Ibaraki 305-0801, Japan}
\author{N.~Terunuma}
\affiliation{KEK: High Energy Accelerator Research Organization, 1-1 Oho, Tsukuba, Ibaraki 305-0801, Japan}
\affiliation{SOKENDAI: The Graduate University for Advanced Studies, 1-1, Oho, Tsukuba, Ibaraki 305-0801, Japan}
\author{J.~Urakawa}
\affiliation{KEK: High Energy Accelerator Research Organization, 1-1 Oho, Tsukuba, Ibaraki 305-0801, Japan}

\begin{abstract}
Investigation of coherent Smith-Purcell Radiation (SPR) spectral characteristics was performed both experimentally and by numerical simulation. The measurement of SPR spectral line shapes of different diffraction orders was carried out at KEK LUCX facility. A pair of room-temperature Schottky barrier diode (SBD) detectors with sensitivity bands of $60-90$~GHz and $320-460$~GHz was used in the measurements. Reasonable agreement of experimental results and simulations performed with CST Studio Suite justifies the use of different narrow-band SBD detectors to investigate different SPR diffraction orders. It was shown that monochromaticity of the SPR spectral lines increases with diffraction order. The comparison of coherent transition radiation and coherent SPR intensities in sub-THz frequency range showed that the brightnesses of both radiation mechanisms were comparable. A fine tuning feasibility of the SPR spectral lines is discussed. 
\end{abstract}


\maketitle

\section{Introduction}
Intense THz radiation is widely used for different applications including THz diffraction and spectroscopy~\cite{1,2}. The interest appears due to the fact that THz radiation is non-ionizing, which prevents destruction of a sample and enables investigation of living cells without radiation damage. Furthermore, most of the molecules oscillate at THz frequencies providing distinct spectral signatures of different materials when the radiation propagates through the sample~\cite{3}. 

These days THz radiation can be produced by tabletop type thermal or laser mixer based generators~\cite{4,5,6}. However, the power of these sources is rather low. The state-of-the-art methodology for high power THz generation is based on particle accelerators, e.g. ultra-short pulse compact accelerators~\cite{7} or free-electron lasers~\cite{7a}. Nevertheless, an optimal mechanism for THz production is still under consideration. The problem of designing of a cost-effective, compact, adjustable THz source with short pulse duration has to be resolved.

There are several approaches based on electron beam technologies (see, for instance,~\cite{8,9,10,11}) proposed to design such a source. In most cases an adjustable monochromator needs to be foreseen to achieve a narrow-band THz output, but tunable in a broad spectral range. However, the usage of any kind of diffractometer or bandpass filter reduces the transmitted power and may introduce an undesirable spectra distortion in virtue of diffraction effects. In this respect, a THz source based on Smith-Purcell radiation (SPR) mechanism is promising, because SPR appearing when a charged particle moves above and parallel to a grating is a resonant process with spectral lines defined by the well-known dispersion relation:
\begin{equation}
\label{eq1}
\lambda _{k} = \frac{{d}}{{k}}\left(\frac{1}{\beta}-\cos\theta\right).
\end{equation}

\noindent
Here $\lambda _{k}$ is the wavelength of the resonance order $k$, $d$ is the grating period, $\beta$ is the particle velocity in the speed of light units, and $\theta$ is the observation angle.
 
The use of coherent SPR generated by short electron bunches (or by a train of bunches) as the basis of THz radiation sources was proposed by authors of the Refs.~\cite{12,12a,12b,13}. Coherent radiation emission occurs when the bunch length is comparable to or shorter than the radiation wavelength resulting in the SPR intensity being determined by the square number of electrons per bunch. The spectral-angular distribution of coherent SPR produced by a strip grating with a finite number of periods $N$ can be written as:
\begin{equation}
\label{eq2}
\begin{array}{l}
\displaystyle \frac{{d^{2}W}}{{d\nu d\Omega} } =
\displaystyle  \frac{{d^{2}W_{0}} }{{d\nu d\Omega
}}{\frac{{\sin^{2}\left[ {N \phi} \right]}}{{\sin^{2}\left[
{\phi} \right]}}} \left(N_{e}+N_{e}\left(N_{e}-1\right) F\right). 
\end{array}
\end{equation}

\noindent
Here $\phi = d{{{\pi \nu}\over{c}}}\left( {\beta ^{ - 1} - \cos\theta} \right)$, $d^{2}W_{0} /d\nu d\Omega $ is the spectral-angular distribution of the radiation from a single electron, $\nu$ is the radiation frequency $\left( {\nu = c/\lambda}  \right)$, $N_{e} $ is the bunch population, and $F$ is the bunch form-factor~\cite{14}. According to Eq.~(\ref{eq2}) extension of the number of periods to infinity $N \to \infty $ results in Dirac's delta function that defines dispersion relation in Eq.(\ref{eq1}). Also intensity $d^{2}W_{0} /d\nu d\Omega $ decreases if the relation $ \gamma \lambda_{k} \le b $ is not fulfilled~\cite{15}, where $\gamma$ is the the electron Lorenz-factor, $b$ is the grating width, which is the size along the direction perpendicular to electron momentum.

\section{Principles}
One of the most important characteristics of any compact accelerator based THz source is its monochromaticity. From Eq.(\ref{eq2}) the monochromaticity of the SPR spectral line is ($\Delta\lambda_k/\lambda_k \propto 1/{kN}$) for the finite length grating $Nd$. Using full width at half maximum (FWHM) as an absolute spectral line width, the monochromaticity is defined as:
\begin{equation}
\label{eq3}
\frac{\Delta\lambda_k}{\lambda_k} = \frac{\Delta\nu_k}{\nu_k} = \frac{0.89}{kN},
\end{equation}
stating the fact that higher SPR orders are more monochromatic than the fundamental one. 

As it was shown in~\cite{14} the value of the SPR line width measured by a detector placed in the so-called prewave zone becomes broader than the estimation given by Eq.~(\ref{eq3}). If the grating-to-detector distance is $L$, then the far-field zone (or wave zone) condition is determined by~\cite{15}:
\begin{equation}
\label{eq4}
L \gg L_{ff} = kN^{2}d\left(1+\cos\theta\right).
\end{equation}
Contrariwise when $L\le L_{ff}$ the condition complies with the prewave zone and the simplest way to avoid spectral line broadening is to use focusing optical elements (e.g. lenses or focusing mirrors) in front of the detector~\cite{16,17}. However, such an effect does not influence the monochromaticity $ \Delta \lambda /\lambda $.  The expression for monochromaticity in this case can be derived directly from the Eq.~(\ref{eq1}) as:
\begin{equation}
\label{eq5}
\frac{\Delta\lambda}{\lambda}= \frac{\Delta\nu}{\nu} =\frac{\sin\theta}{1/\beta-\cos\theta}\Delta \theta.
\end{equation}

Equation~(\ref{eq5}) determines the monochromaticity of radiation generated from an infinite grating ($N \to \infty $) and measured with a finite detector aperture $\Delta \theta$. Nevertheless, Eq.~(\ref{eq5}) allows to estimate the broadening of SPR spectral line due to finite aperture of the detector used during the experiment. In addition, any spectrometer possesses its own intrinsic resolution; hence, the spectral line shape measurements always include systematic distortion. Assuming that the initial line shape $\Delta\lambda^{SPR}$ and spectrometer resolution $\Delta\lambda^{int}$ can be approximated by a Gaussian distributions, one can use the following expression for the FWHM value of the measured line width:
\begin{equation}
\label{eq6}
\Delta \lambda_k = \sqrt {\left(\Delta\lambda^{SPR}_k\right)^2+\left(\Delta\lambda^{int}_k\right)^2}.
\end{equation}

According to Eq.~(\ref{eq3}) transition to the higher diffraction orders ($k>1$) allows to narrow the coherent SPR spectral line $\Delta\lambda^{SPR}$ significantly if the angular acceptance forming the radiation beam is chosen to be as small as practically possible. The main objective of this paper is to show a possibility to generate SPR beams with monochromaticity better than $1-2\%$ choosing the higher diffraction order $k>1$ even for $N$ of about $10$.

To the best of our knowledge, the authors of the Ref.~\cite{18} have observed coherent SPR for the first time. They have measured the SPR spectrum generated by the $42$~MeV electrons from a $4$~mm period grating with $N=20$ periods using the grating-type spectrometer with a focusing mirror. The observed SPR line at $\theta = 70^{\circ}$ for $k=1$ had the wavelength of $\lambda_1=2.68$~mm with FWHM $\Delta\lambda = 0.21$~mm. Taking into account the bandwidth of the spectrometer in use, $\Delta\lambda^{int}=0.06$~mm, it is possible to conclude that the observed line broadening was caused by both above-mentioned reasons (see Eqs.~(\ref{eq5}) and (\ref{eq3})). In~\cite{19} the SPR spectrum generated by the $2.3$~MeV electrons was measured with a Fabry-Perot interferometer. The grating with $2.5$~mm period and effective length of $20$~mm was used. From the experimental results obtained for the angle $\theta = 90^{\circ}$ and the first diffraction order ($k=1$, $\lambda _1 = 2.5$~mm) presented in Fig.~$1$~\cite{19}, one can estimate the monochromaticity of the measured spectral line as $\Delta\lambda _1/\lambda_1=15\%$. Such a large value for interferometer can be explained by both factors (\ref{eq5}), (\ref{eq3}) and, probably, by the prewave zone effect (in addition, an extra uncertainty is coming from the fact that the geometry of the experiment was not presented by the authors). In~\cite{20} the coherent SPR characteristics from the finite length and width grating were considered. In order to avoid losses of the radiated power in comparison with the conventional case (i.e. when transverse grating size tends to infinity), the requirements for the grating width were formulated and the power spectrum as well as the radiated energy using different models were calculated. However, the simulation of the SPR line widths was not considered.

\section{Experiment} \label{Experiment}
\subsection{Methods and techniques}
\begin{figure}[ht]
\includegraphics[width=86mm]{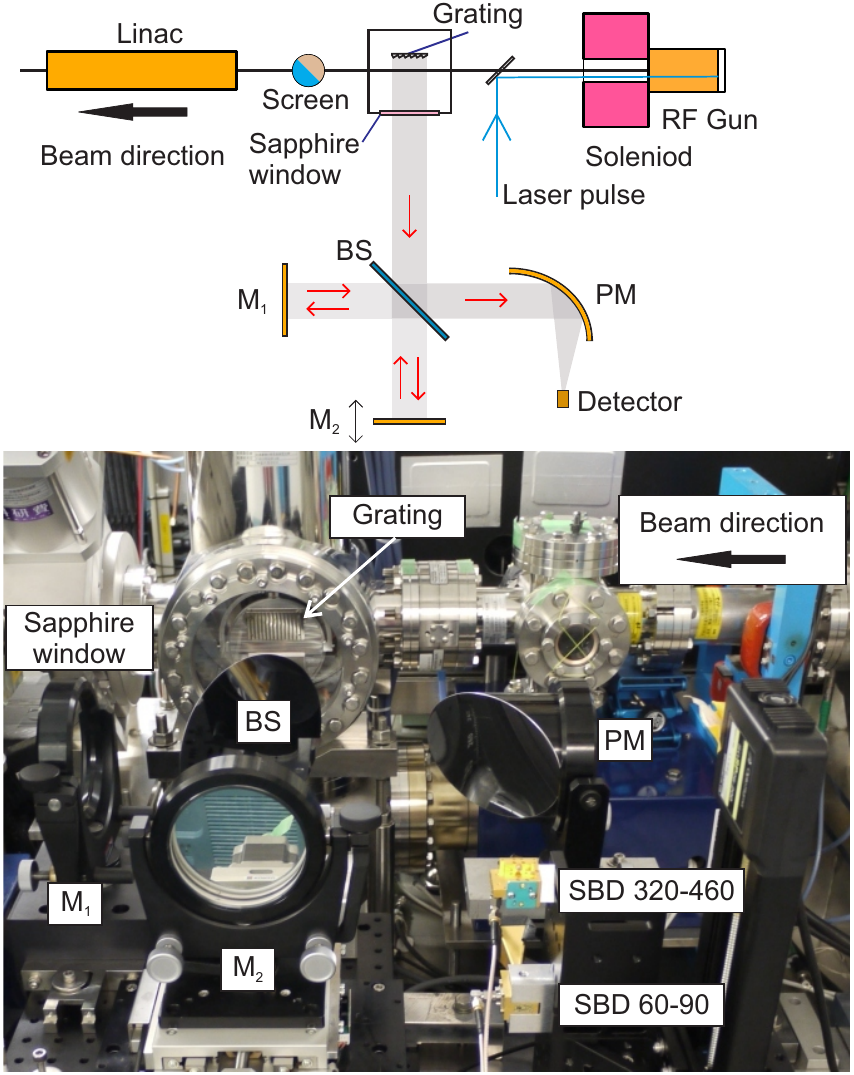}
\caption{Top: experimental schematics. Bottom: photograph of the experimental station. Abbreviations: M$_1$ - fixed interferometer mirror, M$_2$ - movable interferometer mirror, BS - splitter, PM - off-axis parabolic mirror.}
\label{SPR_scheme}
\end{figure}

The experiment was carried out at the KEK LUCX facility. Schematic diagram and the photograph of the experimental setup are shown in Fig.~\ref{SPR_scheme}.

Short electron bunches were generated in the RF gun via Cs$_{2}$Te photocathode illumination by the femtosecond laser pulses with wavelength of $266$~nm. Then electron bunches were accelerated to the energy of $8$~MeV in the $3.6$~cell RF gun. The experiment was conducted with electron beam parameters given in Table~\ref{T1}. The transverse shape of the electron beam in experimental area was measured using a scintillating screen, which was located $\sim 400$~mm downstream the vacuum chamber with installed gratings. The longitudinal electron bunch profile measurements were well described in~\cite{21}.

\begin{table}[h!]
\caption{KEK: LUCX, beam parameters at the RF gun section.}
\label{T1}
\begin{ruledtabular}
\begin{tabular}{l l}
  Parameter & Value\\
  \hline
  Beam energy & $8$~MeV\\
  Charge per bunch & $25$~pC\\
  Bunch rms length & $0.5$~ps\\
  Transverse rms size & $230\times230~\mu$m\\
  Repetition rate & $3.13$~bunch/s\\
  Normalized emittance, typ. & $1.5\times1.5$~mm mrad\\
\end{tabular}
\end{ruledtabular}
\end{table}

The vacuum chamber for experimental investigation was installed after the RF gun. The chamber vacuum window was made of the $12$~mm thick 2$^{\circ}$ wedged sapphire and mounted into an ICF-203 flange, which provided an effective aperture of $145$~mm. A $5$-axis manipulator system was integrated on the top of the chamber. It was used for fine adjustment of grating's position in $3$ orthogonal directions and also for the control of the two rotation angles of the grating with respect to the electron beam propagation direction. The mechanics of the manipulator were based on the linear and rotation stages driven by stepping motors with resistive encoders. All motors were remotely controlled by an industrial-grade Oriental Motor CRK-series controllers in the open-loop mode~\cite{22,23}. The positioning accuracies were better than $5~\mu$m and $0.02^{\circ}$ for the linear and rotation stages, respectively. This allowed to control the grating position with respect to the electron beam trajectory. Prior sub-THz radiation properties measurements, the grating was aligned with respect to electron beam using the forward bremsstrahlung appearing due to direct interaction of the electron beam with the grating material. In the case of SPR geometry, the distance between grating and the electron beam was $0.6$~mm. The radiation spectral characteristics were measured by the Michelson interferometer (described in~\cite{23}) installed directly in front of the chamber vacuum window (see Fig.~\ref{SPR_scheme}). The main interferometer optical axis coincided with the direction perpendicular to the electron beam propagation and corresponded to the observation angle $\theta =\pi/2$. Two Schottky barrier diode detectors (SBD) with different regions of spectral sensitivity $60-90$~GHz and $320-460$~GHz were used in the experiment. Detailed detectors' parameter list is shown in Table~\ref{T2}. The electron beam parameters (bunch length $\approx 0.5$~ps) ensured coherent radiation emission within the measurement spectral regions. In this case the radiation intensity was scaled by the bunch population factor $N_e \approx 1.56\cdot10^8$ and the detectors described above could be used for radiation intensity measurements. 

\begin{table}[h!]
\caption{Detector parameters.}
\label{T2}
\begin{ruledtabular}
\begin{tabular}{l l l}
  Parameter & SBD~60-90~GHz & SBD~320-460~GHz\\
  \hline
  Frequency range & $60-90$~GHz & $320-460$~GHz\\
  Wavelength range & $3.3-5.0$~mm & $0.94-0.65$~mm\\
  Response time & $\sim250$~ps & sub-ns\\
  Antenna gain & $24$~dB & $25$~dB\\
  Input aperture & $30\times23$~mm & $4\times4$~mm\\
  Video sensitivity & $20$~V/W & $1250$~V/W\\
\end{tabular}
\end{ruledtabular}
\end{table}
\begin{figure}[h!]
\includegraphics[width=86mm]{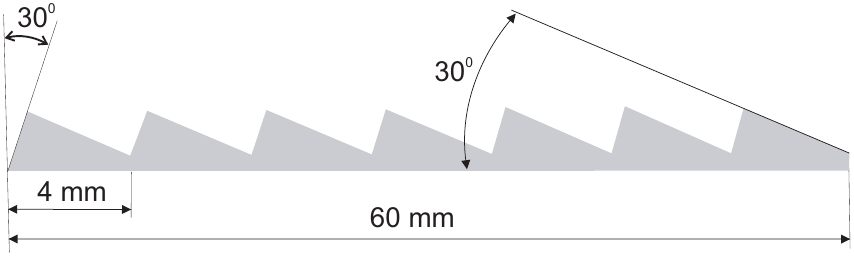}
\caption{SPR grating geometry.}
\label{SPR_target}
\end{figure}

The $60\times30$~mm$^2$ echellete profile grating shown in Fig.~\ref{SPR_target} was placed in the vacuum chamber to generate Smith-Purcell radiation. The opposite side of the grating was flat that allowed to use this surface as coherent transition radiation (TR) source. The grating could be rotated around its vertical axis for TR orientation dependence measurement (co-called $\Theta$-scan, where $\Theta$ is the angle between TR target surface and electron beam direction). During such a scan the interferometer movable mirror was set to the position of zero path difference.

\subsection{Transition radiation characteristics}

\begin{figure}[ht]
\includegraphics[width=86mm]{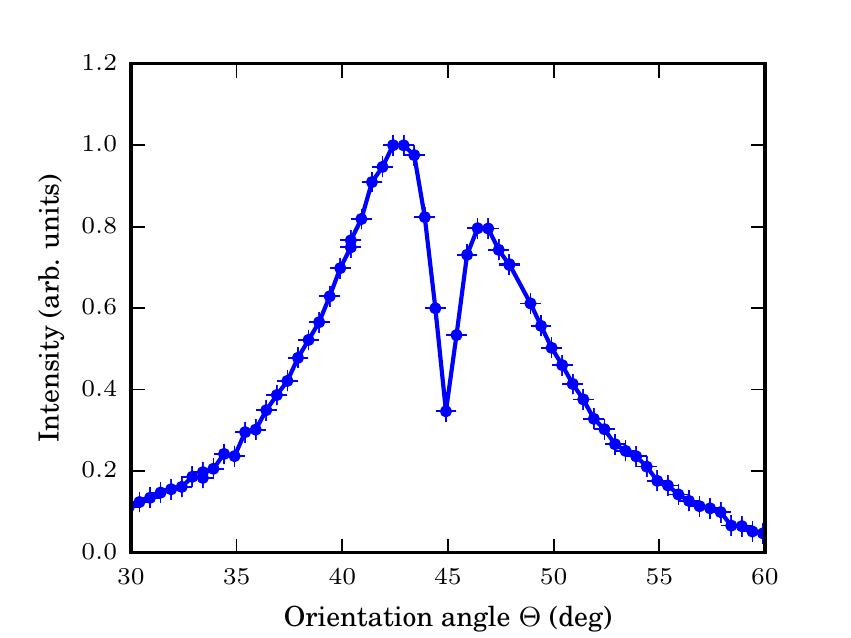}
\caption{TR orientation dependence measured in the frequency range $320-460$~GHz.}
\label{TR_orientation_SBD_320-460}
\end{figure}

From the theory~\cite{14} the TR spectrum emitted by a single electron is supposed to be constant within detector sensitivity bands. Therefore the measured normalized TR spectrum can be used as the entire measurement system spectral efficiency, including spectral transmission efficiency of the vacuum window, detector wavelength efficiency, splitter efficiency, reflection characteristics of the mirrors and air absorption. The typical TR orientation dependence obtained by rotation of the TR target at the angle $\Theta$ and measured with $320-460$~GHz SBD is shown in Fig.~\ref{TR_orientation_SBD_320-460}.

\begin{figure}[ht]
\includegraphics[width=86mm]{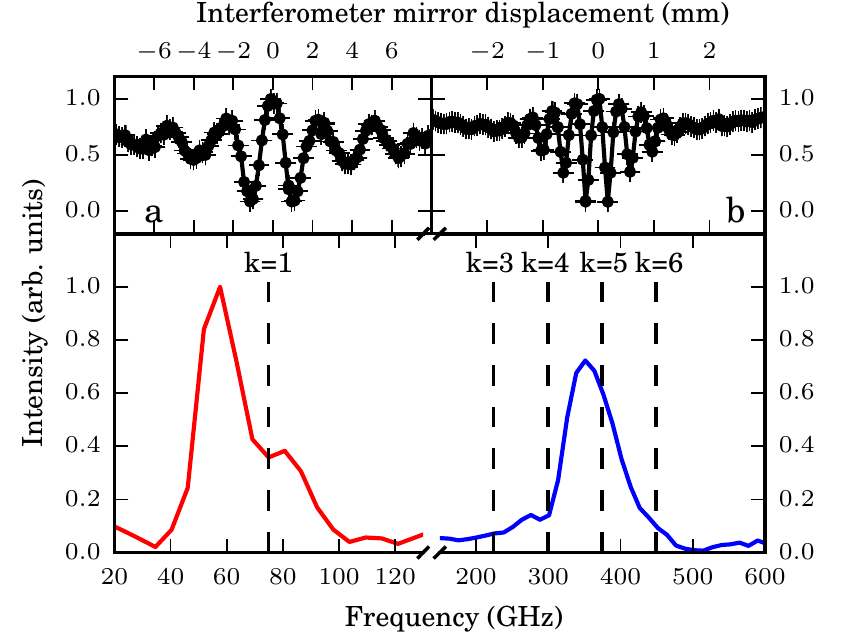}
\caption{TR spectral measurements results. $a$ and $b$ are interferograms measured in the range of $60-90$~GHz and $320-460$~GHz respectively. Bottom plots: reconstructed spectra.}
\label{TR_spectra}
\end{figure}

Interferograms measured using Michelson interferometer at the maxima of orientation dependencies with both detectors are shown in Fig.~\ref{TR_spectra}$a$ and Fig.~\ref{TR_spectra}$b$.

Fourier transform algorithm was used for spectral reconstruction~\cite{24}. Two normalized TR spectra measured in a range of $60-90$~GHz and $320-460$~GHz are shown in Fig.~\ref{TR_spectra} (bottom part). Subsequently, these spectra were used as the entire measuring system spectral efficiency for the SPR spectra renormalization. Vertical dashed lines correspond to the SPR diffraction orders $k$ calculated using the dispersion relation Eq.~(\ref{eq1}) for experimental grating parameters and observation angle $\theta =\pi/2$. The resonances $k=1$ and $k=5$ were within the detector sensitivity bands.

\subsection{SPR characteristics}

\begin{figure}[ht]
\includegraphics[width=86mm]{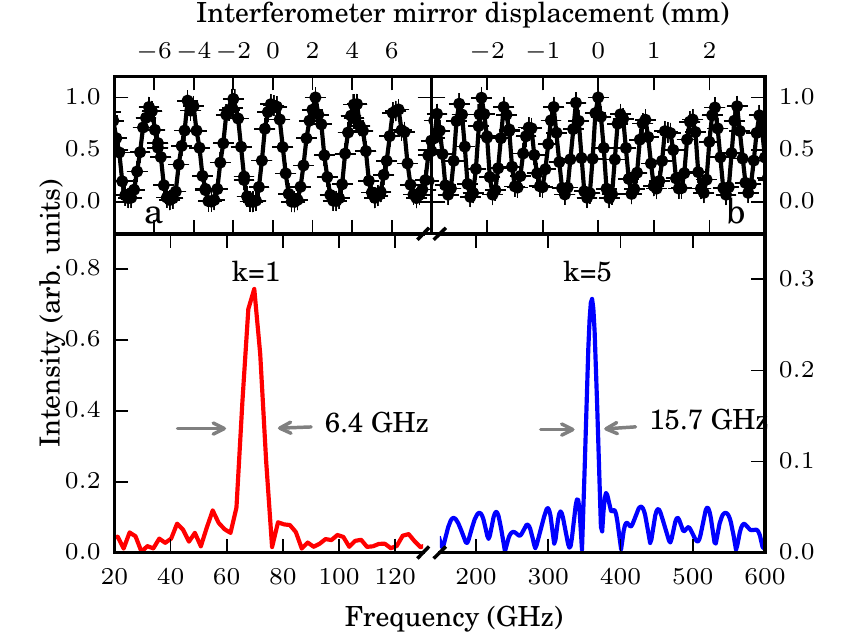}
\caption{SPR spectral measurements results. $a$ and $b$ are interferograms measured in the range of $60-90$~GHz and $320-460$~GHz respectively. Bottom plots: reconstructed spectra.}
\label{SPR_spectra}
\end{figure}

Measured SPR interferograms are shown in Fig.~\ref{SPR_spectra}.
Since there are different definitions of Fourier spectrometers resolution, it is important to mention that the criterion, which defines $\Delta\lambda^{int}_k$ as FWHM of the apparatus spectral peak from monochromatic source with the wavelength $\lambda_k$ was chosen:
\begin{equation}
\label{eq7}
\frac{\Delta\lambda^{int}_k}{\lambda_k}=1.21\frac{\lambda_k}{2L_{int}},
\end{equation}
where $L_{int}$ is the interferometer maximal optical paths difference from zero position. For Michelson interferometer in case of symmetrical interferogram this value coincides with the full length of the interferogram. Applying this criterion to the interferograms shown in Fig.~\ref{SPR_spectra} (top part) it is possible to find $\Delta\lambda^{int}_1/\lambda_1 = 4.2\%$ for $60-90$~GHz and $\Delta\lambda^{int}_5/\lambda_5 = 2.1\%$ for $320-460$~GHz. The spectra recovered from these interferograms are shown in Fig.~\ref{SPR_spectra} (bottom part). As one can see the spectral peak measured in the range $60-90$~GHz corresponds to $k=1$ and the spectral peak measured in $320-460$~GHz range corresponds to the $5^{\text{th}}$ SPR resonance $k=5$. The peaks' relative line widths are $\Delta\lambda_1/\lambda_1 = 8.8\%$ and $\Delta\lambda_5/\lambda_5=4.3\%$, which are close to the estimated spectral resolution $\Delta\lambda^{int}_k/\lambda_k$ obtained through analysis of the interferometer characteristics. It is obvious that the measured spectra peaks' widths are defined by the interferometer spectral resolution and the real peak widths are narrower. To compare the radiation intensities at $k=1$ and $k=5$ the SPR spectra were normalized by TR spectra. It is also important to notice that the background contribution (both coherent from the accelerator beamline and associated with environmental noise) to the SBD signal was constantly low during the experimental run.

\section{Simulations}
\begin{figure}[ht]
\includegraphics[width=86mm]{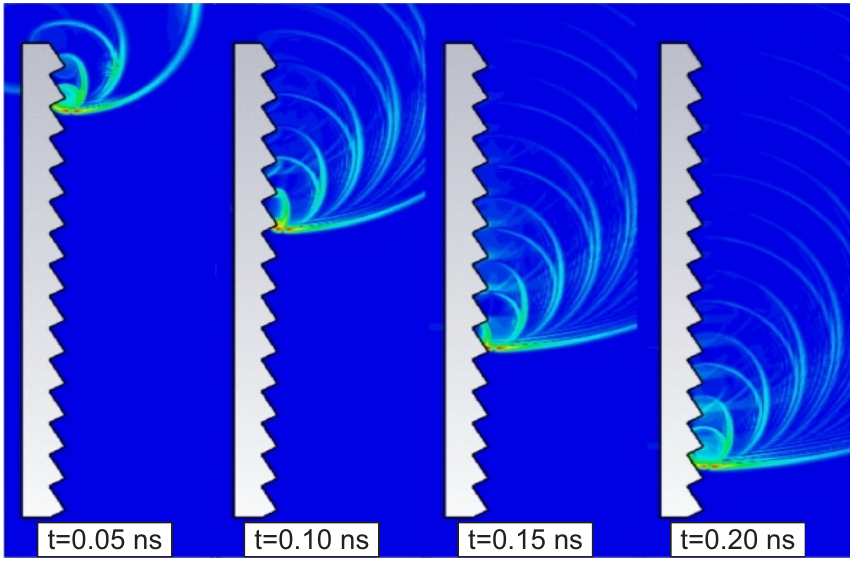}
\caption{Electric field representation of beam propagation through small calculation domain near grating.}
\label{figLecom_4}
\end{figure}

\begin{figure}[ht]
\includegraphics[width=86mm]{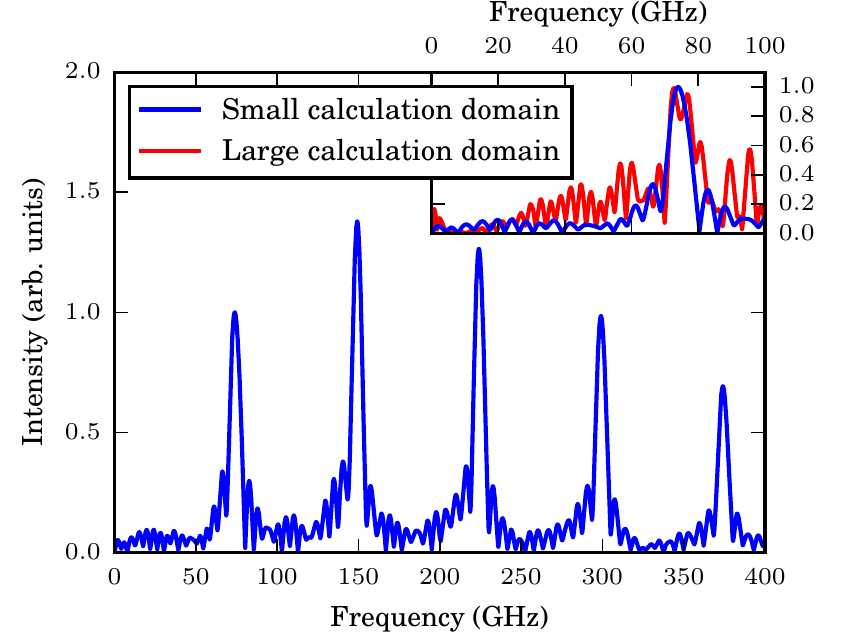}
\caption{Calculated SPR spectrum in small calculation domain. Inset: comparison with SPR spectrum calculated in large calculation domain.}
\label{figLecom_5}
\end{figure}

The spectrum of SPR from the grating, identical to the one used in the experiment (see section~\ref{Experiment}), was simulated using Computer Simulation Technology (CST) Particle In Cell solver~\cite{24a}. According to Eq.~\ref{eq4} even for the first diffraction order the far-field condition is not fulfilled. However, taking into account focusing parabolic optics used in experiment it was assumed that prewave zone effect was minimized~\cite{16,17}. The distance from the grating to the electric field probe, located perpendicularly to the grating surface, was equal to $L = 500$~mm. $L$ was also the distance from the grating to the movable mirror (M$_2$) in the interferometer (see Fig.~\ref{SPR_scheme}). The simulations were carried out using two calculation domains in order to show the influence of prewave zone effect for the first diffraction order of radiation. The main calculation was done for the frequency range $0-400$~GHz using so-called ``small calculation domain'' that assumed calculation of the electric fields in the calculation domain with the size $16\times32\times70$~mm$^3$, to enclose entire 3D grating, and the following field propagation based on CST far-field monitor. It was confirmed that the spectrum calculated by the monitor was not sensitive to increase of the domain size ($60\%$ increase was considered), and, therefore, the smaller domain was chosen to reduce calculation time~\cite{24b}. In the case of the small domain the prewave zone effect was not taken into account. A so - called ``large calculation domain'' that covered the whole area of the radiation propagation was used for the frequencies up to $100$~GHz only because of simulation time limitations. In this domain the prewave zone effect was taken into account due to the fact that $L_{ff}=900$~mm. Beam propagation and electric field distribution in the 2D cross-section of the small calculation domain at four consecutive time steps are shown in Fig.~\ref{figLecom_4}.
The electron beam parameters used in the simulation are shown in Table~\ref{T1}. The spectrum of SPR was obtained by recording the dominant component of the electric field at the probe location as a function of time and, then, by performing Fourier transform. This procedure remained the same for both calculation domains, the only difference was that in the case of the small domain the electric field values at the border were extrapolated to the probe location, and for the large domain they were recorded at the probe without extrapolation. Five diffraction orders in the SPR spectrum calculated for the far field are shown in Fig.~\ref{figLecom_5}.


The influence of the prewave zone effect was investigated by comparison of radiation spectra obtained using both calculation domains at frequencies $\nu_1= 0 - 100~$GHz. The inset of Fig.~\ref{figLecom_5} shows the spectra calculated using the two methods: first - when the electric field values at the border of the small domain were extrapolated using the far field monitor (blue curve), and the second - when the electric field values were recorded at the probe in the large domain without extrapolation (red curve). For the blue curve $\Delta \lambda_1/\lambda_1=8.7\%$, for the red curve $\Delta \lambda_1/\lambda_1=9.3\%$, and the theoretical value calculated using Eq.~(\ref{eq3}) is $\Delta \lambda_1/\lambda_1=5.9\%$. Increased width of the spectrum for the second method agrees well with the fact that the prewave zone affects the spectrum. In the case of $1^{\text{st}}$ order the broadening effect should be more pronounced and it should be always taken into account. Noise in the red curve is most likely caused by a low signal to numerical noise ratio for the large domain due to the probe being located at the large distance $L\gg\lambda$ from the grating. Table \ref{T3} summarizes the comparison of the SPR line widths simulated using CST studio suite, Eq. (\ref{eq3}) and measured SPR relative line widths. The $5^{\text{th}}$ diffraction order in Fig.~\ref{figLecom_5} corresponds to the far field. Monochromaticity of this peak is $\Delta \lambda_5/\lambda_5=1.6\%$ and the theoretical value is $\Delta \lambda_5/\lambda_5=1.2\%$. The monochromaticities of the $1^{\text{st}}$ and $5^{\text{th}}$ diffraction orders in the far field spectrum are $1.5$ and $1.3$ times larger than the corresponding theoretical values. Nevertheless, the spectral line of the $5^{\text{th}}$ order is narrower than the line of the $1^{\text{st}}$ order, which agrees well with the theoretical predictions.

\begin{table}[h!]
\caption{Comparison of the SPR spectral line widths: CST simulation, theoretical and measured values.}
\label{T3}
\begin{ruledtabular}
\begin{tabular}{l l l l}
  k & Theory & Simulation & Measurements\\
  \hline
  1 & $5.9\%$ & $8.7\%$ & $8.8\%$ \\
  2 & $3.0\%$ & $3.3\%$ & $-$ \\
  3 & $2.0\%$ & $2.6\%$ & $-$ \\
  4 & $1.5\%$ & $2.0\%$ & $-$ \\
  5 & $1.2\%$  & $1.6\%$ & $4.3\%$ \\
\end{tabular}
\end{ruledtabular}
\end{table}

\section{Results and Conclusion}
We have investigated coherent SPR spectral characteristics both experimentally and by numerical simulation. Our experimental apparatus included the Michelson interferometer with SBD detectors placed in the focal plane of the parabolic mirror located behind the interferometer. To measure SPR spectral line width of different orders the required detector spectral range should be rather broad. Room-temperature detectors such as SBD cover half an octave bandwidths, i.e. $60-90$~GHz and $320-460$~GHz in our case. The spectral range and efficiency of each detector were  investigated experimentally by measuring coherent TR spectrum. The measured results in comparison with theoretical simulations showed a possibility to use different SBD detectors to investigate two different SPR spectral lines ($k = 1$ and $k = 5$). 

For experimental parameters ($E_{e} = 8$\, MeV, $N = 15$, $d = 4$\,mm) the natural (FWHM) spectral line widths according to Eq.~(\ref{eq7}) are equal to $\Delta\lambda^{int}_1/\lambda_1 = 4.2\%$ and $\Delta\lambda^{int}_5/\lambda_5 = 2.1\%$. Thus it is possible to estimate absolute (FWHM) SPR line widths as $\Delta\lambda_1/\lambda_1 = 7.7\%$ at $k=1$ and $\Delta\lambda_5/\lambda_5 = 3.7\%$ at $k=5$, using Eq.~(\ref{eq6}) and taking into account the interferometer resolution. It is in a reasonable agreement with the simulated values (see Tab.~\ref{T3}). Small discrepancy at high frequencies related to the fact that the limited detectors apertures result in increased angular acceptance as:
\[
\Delta \theta = \frac{{\sqrt {A_{eff}} } }{{2f_{par}} },
\]
where $A_{eff} $ is the effective area of the detector and $f_{par}=152\,$mm is the parabolic mirror focal distance ($PM$ at the Fig.~\ref{SPR_scheme}). According to ~\cite{25,26} the effective area of a detector with horn antenna is defined as:
\[
A_{eff} = \frac{{\lambda ^{2}}}{{4\pi} }G,
\]
where $G$ is the antenna gain. For the SBD detectors' parameters shown in Table~\ref{T2} and constant wavelengths $\lambda _{1} = 4\,$mm and $\lambda _{5} = 0.8\,$mm the angular acceptance of $320-460$~GHz SBD is more than $5$ times larger than that of $60-90$~GHz SBD. 

It was shown that the monochromaticity of the SPR spectral lines increases with diffraction order $k$. If angular aperture is defined by:
\[
\Delta \theta < \frac{\tan\left(\theta/2\right)}{{kN}},
\]
the monochromaticity will be determined by the diffraction order $k$ and the number of grating periods $N$. The relative line width smaller than $1\%$ can be achieved.

We have compared intensities of coherent TR and coherent SPR in GHz frequency range measured in identical conditions and showed that the brightness (energy per unit solid angle and per unit frequency range) is practically comparable for both mechanisms. In order to compare coherent TR and SPR intensities at lower frequencies (as presented in \cite{27}) further experimental work is required. The simulation results showed that the intensity of the fifth order spectral line is about two times smaller in comparison with the line intensities of the first and the second orders at $\theta =\pi/2$. Different energy distribution between diffraction orders can be achieved by selecting either different grating parameters or observation angle $\theta$, which is usually limited by the experimental geometry. Also some adjustment of the SPR spectral line can be achieved by the grating tilt angle $\theta _{gr} \ll 1$~\cite{28} with the line shift described by the following relation:
\[
\Delta \lambda = - \frac{\lambda\sin\theta}{1-\cos\theta}\Delta\theta _{gr}.
\]
Another important approach was demonstrated by authors of the Ref.~\cite{29} where a train of short bunches with the fixed spacing was used to generate quasi-monochromatic radiation using the emission mechanism characterised by a continuous spectrum (TR, for instance). One can expect that a train consisting of $N_b$ bunches may be used for SPR monochromaticity improvement if condition $N_b > N$ is fulfilled.

\begin{acknowledgments}
The work was performed by international collaboration AGTaX. Authors would like to thank S.~Araki and M.~Fukuda for valuable help, useful discussion and support of the LUCX accelerator operation and maintenance. This work was supported by the Photon and Quantum Basic Research Coordinated Development Program from the Ministry of Education, Culture, Sport, Science and Technology, Japan, JSPS KAKENHI: $23226020$ and $24654076$, JSPS and RFBR under the Japan-Russia Research Cooperative Program (no.~15-52-50028 YaF\_a), the Leverhulme Trust through the International Network Grant (IN$-2015-012$) and the program "Nauka" of the Russian Ministry of Education and Science within the grant no. 3958 and the European Union Horizon 2020 research and innovation programme under the Marie Sklodowska-Curie grant agreement No 655179.
\end{acknowledgments}

\end{document}